\documentclass[a4paper]{jpconf}
\usepackage{graphicx}
\usepackage{color}

\begin{document}
\title{Stochastic background of gravitational waves from cosmological sources}

\author{Chiara Caprini}

\address{CNRS, URA 2306 and CEA, IPhT, F-91191 Gif-sur-Yvette, France}

\ead{chiara.caprini@cea.fr}

\begin{abstract}
Gravitational waves (GW) can constitute a unique probe of the primordial universe. In many cases, the characteristic frequency of the emitted GW is directly related to the energy scale at which the GW source is operating in the early universe. Consequently, different GW detectors can probe different energy scales in the evolution of the universe. After a general introduction on the properties of a GW stochastic background of primordial origin, some examples of cosmological sources are presented, which may lead to observable GW signals. 
\end{abstract}

\section{Introduction}

Several processes operating in the very early universe can act as sources of a stochastic background of GW. GW of cosmological origin form a fossil radiation: expansion prevents them from coming in thermal equilibrium with the other components of the universe, because of the weakness of the gravitational interaction. Therefore, a stochastic background of GW can contain in its amplitude, spectral shape and frequency range much information on the nature of the source that generated it. The detection of such a fossil background would thus have a profound impact on our knowledge of the early universe, possibly analogous to the impact of the measurement of the Cosmic Microwave Background (CMB) and of its temperature fluctuations. One difference is that GW are out of thermal equilibrium since the Planck scale, while photons decoupled at a temperature of around 0.3 eV. Therefore, relic GW are a potential source of information on the state of the universe at high energies, even higher than those that can be reached in terrestrial laboratories. 

\section{Overview of the GW signal from primordial sources}

In a cosmological context, GW may be represented by a tensor perturbation $h_{ij}$ ($i, j = 1, 2, 3$) of the 
FRW metric $ds^2 = -dt^2 + a^2(t) \, (\delta_{ij} + h_{ij}) \, dx^i dx^j$  
which is transverse and traceless
$\partial_i h_{ij} = h_{ii} = 0$ (we assume flat spatial sections, $a(t)$ is the scale factor and repeated indices are summed). The transverse-traceless condition leaves only two independent physical degrees of freedom. In Fourier space, their linearized equation of motion is
$\ddot{h}_{ij}(\mathbf{k}, t) + 3 H \, \dot{h}_{ij}(\mathbf{k}, t) + 
{k}^2 \, h_{ij}(\mathbf{k}, t) = 16\pi G \, \Pi_{ij}^{(TT)}(\mathbf{k}, t)$, where $G$ is the Newton constant, $H$ is the Hubble rate, a dot denotes derivative with respect to $t$, $k$ is the physical wavenumber and 
$\Pi_{ij}^{(TT)}$ is the transverse-traceless part of the anisotropic stress $\Pi_{ij}$. The latter is given 
by $a^2 \, \Pi_{ij} = T_{ij} - p \, a^2 \, (\delta_{ij} + h_{ij})$, where $T_{ij}$ is the energy-momentum tensor and $p$ the background pressure. Processes that give rise to a non-zero tensor anisotropic stress in the early universe can directly source GW: as we will see, common examples of sources are electromagnetic fields, a scalar field with spatial gradients in its distribution, or the presence of velocity perturbations in the early universe fluid. 

Furthermore, GW can be generated also in the absence of tensor anisotropic stresses, by the amplification of vacuum fluctuations during inflation \cite{inflation}. In this case, the equation of motion for the canonically normalized free field $v_{\pm}=M_{Pl}\,a\,h_{\pm}/\sqrt{2}$ is a homogeneous wave equation, and reads $v''_\pm+a^2(k^2-2H^2)v_\pm=0$ (where a prime denotes derivative with respect to conformal time, $\pm$ are the two polarizations of the tensor modes, $M_{Pl}$ is the reduced Planck mass, and we have set $a''/a\simeq 2a^2H^2$ at lowest order in slow roll). From this equation it appears that particle creation occurs for super-horizon modes with $k\ll H$, because of the fast expansion of the background. The tensor modes thus generated outside the horizon remain constant, until they re-enter the horizon during the radiation or the matter dominated eras: afterwards, they propagate as GW, oscillating and being redshifted by the expansion of the universe. 

The energy density of GW today can be written as $\rho_{gw} = \langle \dot{h}_{ij} \, \dot{h}_{ij} \rangle/(32 \pi G) = 
\int \frac{d f}{f} \, \frac{d \rho_{gw}}{d \log f}$, where $f = (k / 2 \pi) (a/ a_0)$ is the present-day GW frequency ($a_0$ denoting the scale factor today) and 
$\langle \rangle$ denotes ensemble average. A cosmological stochastic background is (at least in a first approximation) statistically isotropic, stationary and Gaussian: its main properties are then described by its power spectrum. One defines the spectrum of energy density per logarithmic frequency interval divided by the critical density $\rho_c$ today, $h^2 \, \Omega_{\rm GW}(f) =\frac{h^2}{\rho_c} \, \frac{d \rho_{gw}}{d \log f}\,,$ where $h$ parametrizes the Hubble constant $H_0 = 100\,h\,$km/s/Mpc. 

A causal GW source that operates at sub-Hubble scales at some time $t_*$ after inflation emits GW with a characteristic wave-number $k_*$ that is larger than the Hubble rate $H_*$ at that time: $k_* = H_* / \epsilon_*$ with 
$\epsilon_* \leq 1$ \cite{Allen:1996vm}. The characteristic GW frequency today for a causal process is then given by $f_c = H_* / (2\pi\epsilon_*) (a_* / a_0)$. For GW generated in the radiation era when the plasma temperature is $T_*$, the characteristic frequency today can thus be written as
$f_c \simeq 2.6 \times 10^{-5} \, \mathrm{Hz}  ~~ \epsilon_*^{-1} \, (T_*/1 \, \mathrm{TeV}) \, 
(g_*/100)^{1/6}\,, $
where we have assumed a standard thermal history 
for the evolution of the universe after GW production, and $g_*$ is the number of relativistic degrees of freedom at temperature $T_*$. The parameter $\epsilon_* \leq 1$ depends on the dynamics of the specific GW source under consideration. Modulo the value of this parameter, it is therefore possible to establish a correspondence between the characteristic frequency today $f_c$ of the GW emitted by a causal source, and the epoch at which the source was operating, identified by $T_*$ (c.f. Fig.~\ref{fig:inf}). Note that this correspondence only holds for GW emitted by causal sources, and not for the GW spectrum possibly generated by inflation, which leads to the amplification of super-horizon tensor modes and gives rise to a spectrum which is almost scale invariant over a wide frequency range, see Fig.~\ref{fig:inf}.

As a result of the correspondence between $f_c$ and $T_*$, different GW detectors operating at different frequency ranges can probe different energy scales in the history of the universe. For a first-order phase transition (PT), one may have for instance $\epsilon_* \sim 10^{-3} - 1$ (see e.g. \cite{elisapaper}). In this case, the above equation for $f_c$ shows that GW produced around the EW scale $T_*\simeq 100$ GeV are potentially interesting for detection with eLISA, which operates in the frequency range $10^{-5}\,{\rm Hz}<f<1\,{\rm Hz}$ \cite{Seoane:2013qna}. On the other hand, GW production from the QCD phase transition (QCDPT) at $T_*\simeq 100$ MeV can fall into the frequency range of detection with pulsar timing array (PTA) \cite{Caprini:2010xv}. Earth-based interferometers such as LIGO/Virgo or the future Einstein Telescope (ET) \cite{earth}, operating at a higher frequency range $1\,{\rm Hz}<f<10^3~{\rm Hz}$, can correspondingly probe higher energy scales, up to about $T_*<10^{10}$ GeV. Currently operating and future GW detectors are therefore potentially capable of probing, via the measurement of relic GW, the state of the universe at energy scales which are currently inaccessible by any other mean: those comprised between the scale of inflation and the TeV scale.  

Note that processes operating in the early universe are not the only possible sources of a stochastic background of GW: a contribution is expected also from the incoherent superposition of radiation due to astrophysical sources which cannot be individually resolved. For example, supermassive black hole binaries generate a stochastic background of GW in the PTA frequency band (see e.g.~\cite{Sesana:2008mz}), white dwarf binaries in the eLISA frequency band (see e.g.~\cite{Farmer:2003pa}), neutron star binaries in the frequency band of Earth-based detectors (see e.g.~\cite{Regimbau:2007ed}). Building accurate models of the astrophysical background is very important for two reasons~\cite{Regimbau:2008nj,Meacher:2014aca}. First of all, this background is interesting in itself since it carries information about the properties of the populations of compact objects that generate it. Secondly, it can be a foreground for the cosmological background (and vice versa): therefore, it has to be modeled accurately in order either to be subtracted, or to define the best frequency windows for searching for the cosmological background.

\section{Status of the observations of a GW stochastic background}
\label{sec:status}

\begin{figure}
\includegraphics[width=21pc]{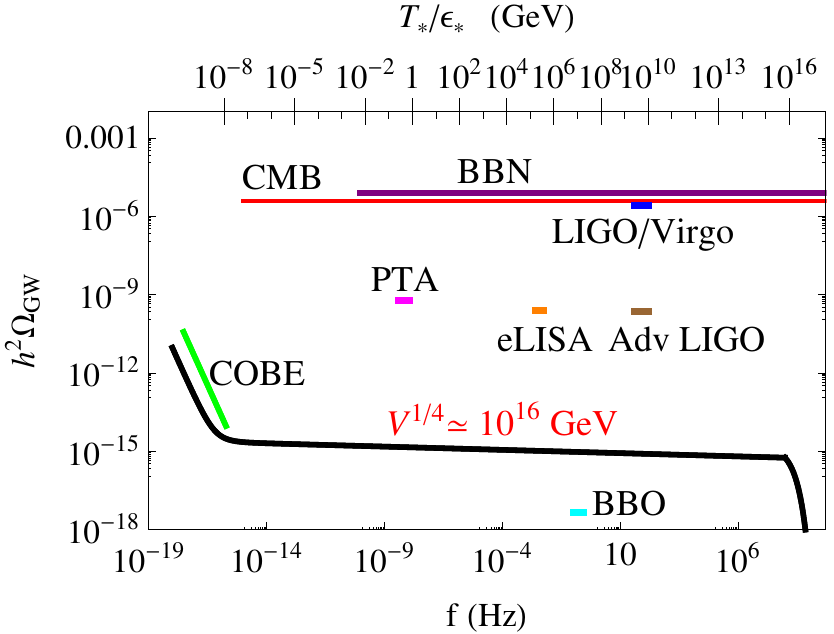}
\hspace*{0.5cm}
\begin{minipage}[b]{2.6in}
\caption{\label{fig:inf} Observational upper bounds on $h^2\Omega_{\rm GW}(f)$, together with the expected sensitivity of some GW detectors (eLISA, Advanced LIGO, BBO - see main text, Sec.~\ref{sec:status}). The upper horizontal axis shows the temperature in the radiation dominated universe at which a causal source can operate in order to emit GW with the characteristic frequency $f$ shown in the lower horizontal axis (modulo the parameter $\epsilon_*$). The black line shows the GW spectrum generated by slow roll inflation at the energy scale $V^{1/4}\simeq 10^{16}$ GeV.}
\end{minipage}
\end{figure}

Current observational upper bounds on the amplitude of the spectrum $h^2\Omega_{\rm GW}(f)$ are shown in Fig.~\ref{fig:inf}. Nucleosynthesis (BBN) and the CMB monopole can be used to establish indirect upper bounds on $h^2\Omega_{\rm GW}(f)$, since they provide a measurement of the total relativistic energy density of the universe respectively at the time of BBN (about 1 MeV), and at photon decoupling (about 0.3 eV). These bounds extend on frequencies higher than the frequency corresponding to the horizon size at BBN and at photon decoupling, and read respectively $h^2\Omega_{\rm GW}^{\rm BBN}(f)\leq 7.8\times 10^{-6}$ for $f_{\rm BBN}>10^{-10}$ Hz \cite{Smith:2006nka}, and $h^2\Omega_{\rm GW}^{\rm CMB}(f)\leq 3.8\times 10^{-6}$ for $f_{\rm CMB}>10^{-16}$ Hz \cite{Smith:2006nka,Henrot-Versille:2014jua}. They are represented in Fig.~\ref{fig:inf} respectively by the purple and red horizontal lines. CMB temperature anisotropies have also been used to set an upper bound on $h^2\Omega_{\rm GW}(f)$, since the presence of tensor perturbations in the FRW metric causes the CMB photons to move on perturbed geodesics and changes their temperature (Sachs Wolfe effect) \cite{Krauss:1992ke}. The bound applies to frequencies corresponding to tensor modes that entered the horizon after the epoch of equality between radiation and matter, and are sub-horizon today: $10^{-18}~{\rm Hz}<f<10^{-16}$ Hz. If the GW spectrum was initially almost scale invariant, as is the case for a tensor spectrum generated during slow roll inflation, the transfer function for tensor modes implies that $\Omega_{\rm GW}(f)$ today decreases as $f^{-2}$ for modes that entered the horizon in the matter era \cite{Krauss:1992ke}. The strongest bound on $h^2 \Omega_{\rm GW}(f)$ is therefore obtained at $f \sim 10^{-16}$ Hz: $h^2 \Omega_{\rm GW} < \mathrm{few} \times 10^{-14}$ \cite{Allen:1996vm}. The bound, at the level given in Ref.~\cite{Allen:1996vm}, is represented by the green line in Fig.~\ref{fig:inf}. 

Other bounds have been established directly by observations with LIGO/Virgo and with PTA: the most recent analyses give respectively, for LIGO/Virgo $\Omega_{\rm GW}(f)\leq 5.6\times 10^{-6}$ at $41.5~{\rm Hz}<f<169.25$ Hz \cite{Aasi:2014zwg}, and for the Parkes PTA $\Omega_{\rm GW}(f)\leq 1.3\times 10^{-9}$ at $f=2.8$ nHz \cite{Shannon:2013wma}. These are represented respectively by the blue and magenta lines in Fig.~\ref{fig:inf} (the value $h=0.67$ has been chosen here \cite{Ade:2013zuv}).

Recently, the BICEP2 experiment detected the presence of B-polarization in the CMB spectrum \cite{Ade:2014xna}. A B-polarization component of primordial origin is generated at photon decoupling from Thomson scattering, if the quadrupole temperature anisotropy of the photons, caused by metric perturbations, has a tensor component. A few months later, new results from the Planck satellite showed that the BICEP2 signal is strongly contaminated by the contribution of dust from the galaxy \cite{Adam:2014bub}. This does not exclude a priori the presence of a contribution from primordial tensor modes of inflationary origin, but it shows that foreground cleaning is absolutely essential. Note that the detection of a stochastic background of GW from inflation would constitute a major discovery: it would imply that gravity can be quantized at the linear level analogously to any canonical quantum field, and would also set the value of the energy scale of inflation in the simplest slow-roll inflationary scenario: $V^{1/4}\simeq 10^{16}$ GeV. The corresponding GW spectrum today in terms of $h^2\Omega_{\rm GW}(f)$ is shown in Fig.~\ref{fig:inf} (black line): here, the tensor to scalar ratio is set to $r=0.2$ and the tensor spectral index to $n_T=-r/8$, we have assumed a standard thermal history after inflation (therefore omitting scenarios as the one described in \cite{BuonBoy}), and we have neglected changes in the number of relativistic degrees of freedom. As explained above, the region $h^2\Omega_{\rm GW}\propto f^{-2}$, for $10^{-18}~{\rm Hz}<f<10^{-16}$ Hz, corresponds to the signal detectable both through temperature anisotropies and B-polarization in the CMB spectrum. On the other hand, the amplitude of the GW spectrum is too low to be directly observable both by currently operating and by forthcoming GW experiments, a part from the conceived space-based observatories BBO and DECIGO \cite{BBODEC}, which are especially designed to reach the inflationary background. The cyan line in Fig.~\ref{fig:inf} represents the maximal sensitivity of the BBO interferometer, taken from Ref.~\cite{Thrane:2013oya} to be at the level of $\Omega_{\rm GW}\simeq 10^{-17}$ at $f\simeq 0.2$ Hz, after one year of observation time.

Figure~\ref{fig:inf} also shows the expected maximal sensitivity of the eLISA interferometer (orange line), taken from Ref.~\cite{AmaroSeoane:2012km} at the level of $h^2\Omega_{\rm GW}\simeq 2.5\times 10^{-10}$ at $f\simeq 3$ mHz, and the expected maximal sensitivity for the correlation of the Hanford-Livingstone pair of Advanced LIGO detectors, taken from Ref.~\cite{Thrane:2013oya} at the level of $\Omega_{\rm GW}\simeq 5 \times 10^{-10}$ at $f\simeq 30$ Hz.

\section{Examples of proposed primordial GW sources}

Besides inflation, there are other possible sources of GW in the early universe which are more promising for detection with future interferometers or PTA. These are all mechanisms that lead to non-zero tensor anisotropic stresses and thereby act as sources of GW. We present in the following some proposals that may lead to observable GW signals (note that this list is non-exhaustive). 

\begin{figure}[h]
\begin{minipage}{18pc}
\includegraphics[width=18pc]{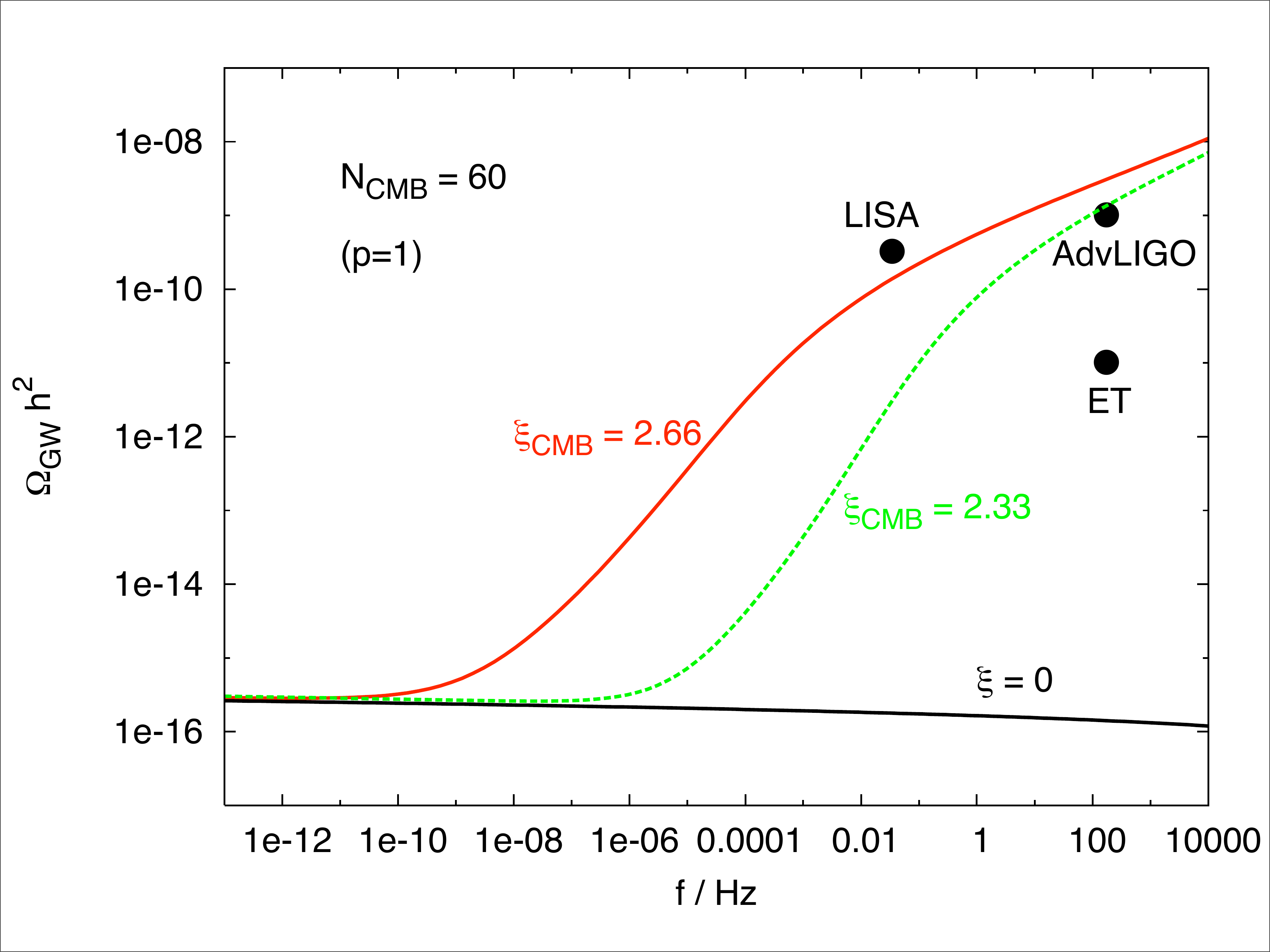}
\caption{\label{fig:PP} Taken from Ref.~\cite{pp2}. The GW spectra (green and red curves) sourced by particle production in a model of pseudoscalar inflation with linear potential ($p=1$, see Ref.~\cite{pp2}) and 60 e-folds of CMB-observable inflation. The spectrum is observable by advanced LIGO if the coupling $\xi_{\rm CMB}=2.33$ (taken at CMB scales). Higher values of the coupling that could possibly lead to a signal observable in LISA are forbidden by CMB non-Gaussianity constraints \cite{meeburgpajer}. $\xi=0$ corresponds to the standard inflationary spectrum.} 
\end{minipage}\hspace{2pc}%
\begin{minipage}{18pc}
\includegraphics[width=18pc]{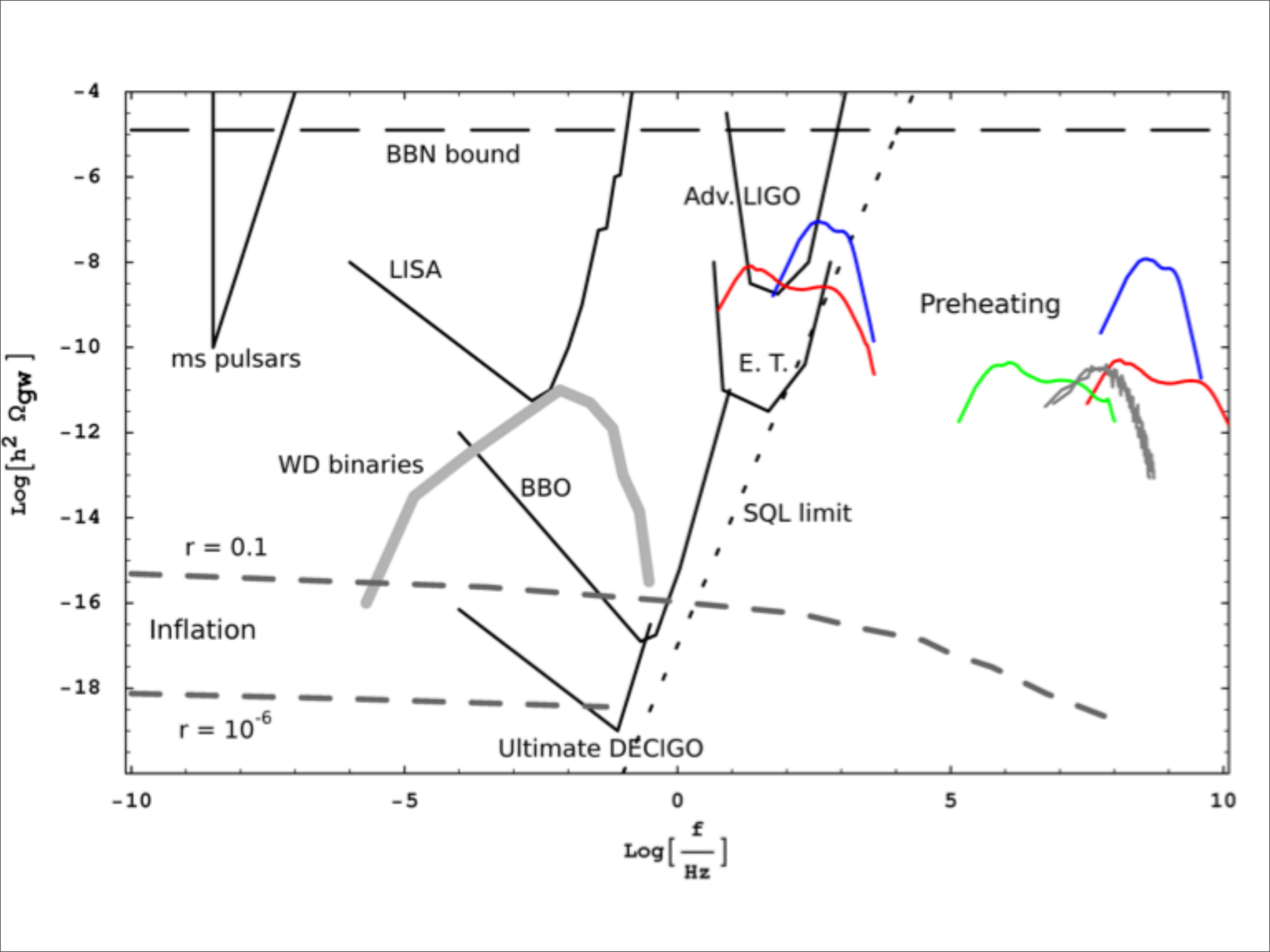}
\caption{\label{fig:pre} Taken from Ref.~\cite{GWHyb}. Examples of GW spectra from tachyonic preheating after hybrid inflation (blue, red and green curves) are shown together with the sensitivity of some GW detectors, the background from extragalactic white dwarf binaries, the inflationary background for two values of the tensor to scalar ratio $r$, and the spectrum from preheating in a  model of chaotic inflation (thin grey curve). The amplitude and peak frequency of the spectra depend on the coupling between the field and the inflaton, its vev, and its initial velocity (c.f.~\cite{GWHyb}).}
\end{minipage} 
\end{figure}

$\bullet$ {\bf Particle production during inflation}: in addition to the amplification of quantum vacuum fluctuations, GW can also be emitted classically during inflation if there are anisotropic stresses. Anisotropic stresses can be generated by particles: as the inflaton rolls down its potential, it provides a time-dependent effective mass to fields coupled to it. If such a field becomes effectively massless during inflation, particles of this field can be produced efficiently in a non-perturbative way \cite{pp1}. The scenario which is the most favorable for detection arises when the inflaton $\phi$ couples to a gauge field through an interaction term of the form $\phi F_{\mu \nu} \tilde{F}_{\mu \nu}$, where $F_{\mu \nu}$ is the field strength and $\tilde{F}_{\mu \nu}$ its dual: this interaction is natural in models where the inflaton field is an axion \cite{pp2}. The gauge field remains massless throughout inflation and its particles can be efficiently produced in a continuous way at horizon crossing, significantly enhancing and polarizing the inflationary GW spectrum up to a level observable by ground-based interferometers \cite{pp2,pp3,Crowder:2012ik}, see Fig.~\ref{fig:PP}. Note however that, even though this enhancement operates far away from the scales probed by the CMB and by Large Scale Structure, it has been recently put forward that these models can be severely constrained by the current bounds on non-gaussianities \cite{Ferreira:2014zia}.

$\bullet$ {\bf Preheating}: GW production can occur also at the end of inflation, when the potential energy density driving inflation is converted into radiation energy density in the course of reheating. In many inflationary models, reheating starts with preheating, an explosive and non-perturbative decay of the inflaton condensate into fluctuations of itself and other fields coupled to it \cite{preheating}. The system then evolves towards thermal equilibrium, but in a highly non-linear and turbulent way. The large field fluctuations lead to non-zero anisotropic stresses and therefore source GW, see e.g. \cite{PreGW1}. Preheating is a causal process and therefore emits GW with a characteristic wavenumber $k_* = H_* / \epsilon_*$ at the time of production, where the Hubble rate during preheating $H_*$ is directly related to the energy scale of inflation. One finds that the inflationary energy scale 
must be smaller than about $10^{11}$ GeV for the peak of the spectrum to fall into the frequency range of ground-based interferometers, and smaller than $10^{7}$ GeV for the peak to fall into the eLISA frequency band. Thus GW from preheating could be observable if inflation occurs at sufficiently low energy: depending on the inflationary scenario, this could be problematic if a component of the BICEP2 signal turns out to be of primordial origin after the complete removal of dust contamination. The peak frequency and the shape of the GW spectrum depend on the inflationary model and on the nature and the couplings of the fields that are produced during preheating. In figure~\ref{fig:pre} we show, as an example, the GW spectrum from tachyonic preheating after hybrid inflation obtained in \cite{GWHyb}. 

$\bullet$ {\bf Cosmic strings}: cosmic strings can be left over by a PT occurring either at the end of inflation or during the thermal evolution of the universe. They are predicted by several high-energy physics models, both in field theory (grand unification, supersymmetry) and in string theory. Once produced, a network of stable cosmic strings evolves towards a self-similar regime characterized by a continuous energy loss: when long string segments cross each other and reconnect, they form smaller cosmic
string loops, which oscillate relativistically and decay away by emitting GW (see e.g. \cite{strings}). Thus the cosmic string energy density
is continually converted into GW via the production of loops: the resulting GW background covers a wide frequency range and can be looked for by different experiments, even simultaneously. Furthermore, cosmic strings also produce GW bursts, emitted by cusps and kinks. The bursts can be looked for either individually, or considering that the incoherent superposition of the burst signals from cusps \cite{Siemens:2006yp} and kinks \cite{Siemens2} also contributes to the stochastic GW background.

\begin{figure}[h]
\begin{minipage}{18pc}
\includegraphics[width=18pc]{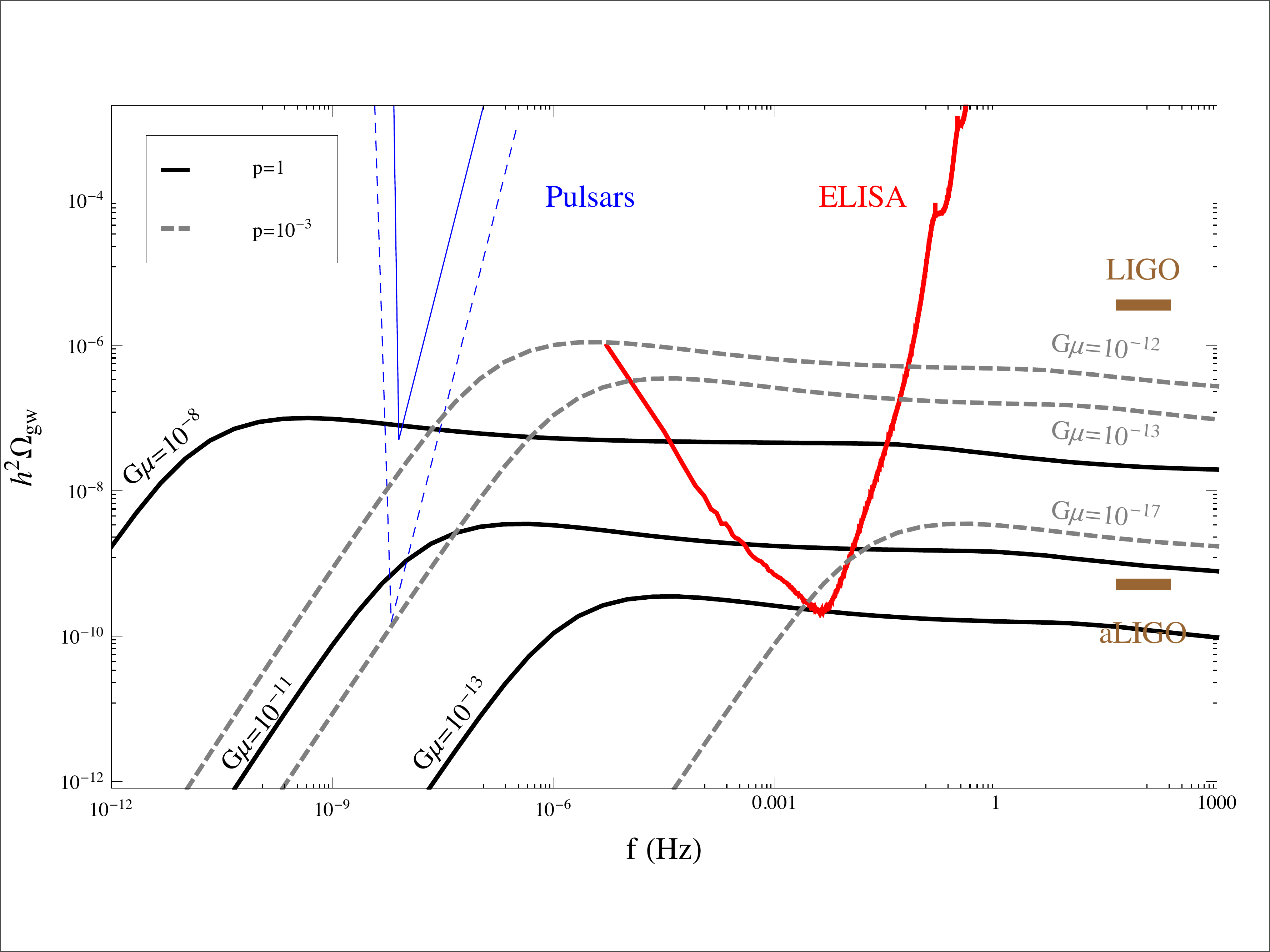}
\caption{\label{fig:strings} Taken from Ref.~\cite{elisapaper}. GW spectra from cosmic strings for different values of the string tension and of the reconnection probability $p$, for large loops, compared to the observational sensitivities of some GW detectors. \vspace*{1.45cm}}
\end{minipage}\hspace{2pc}%
\begin{minipage}{18pc}
\includegraphics[width=18pc]{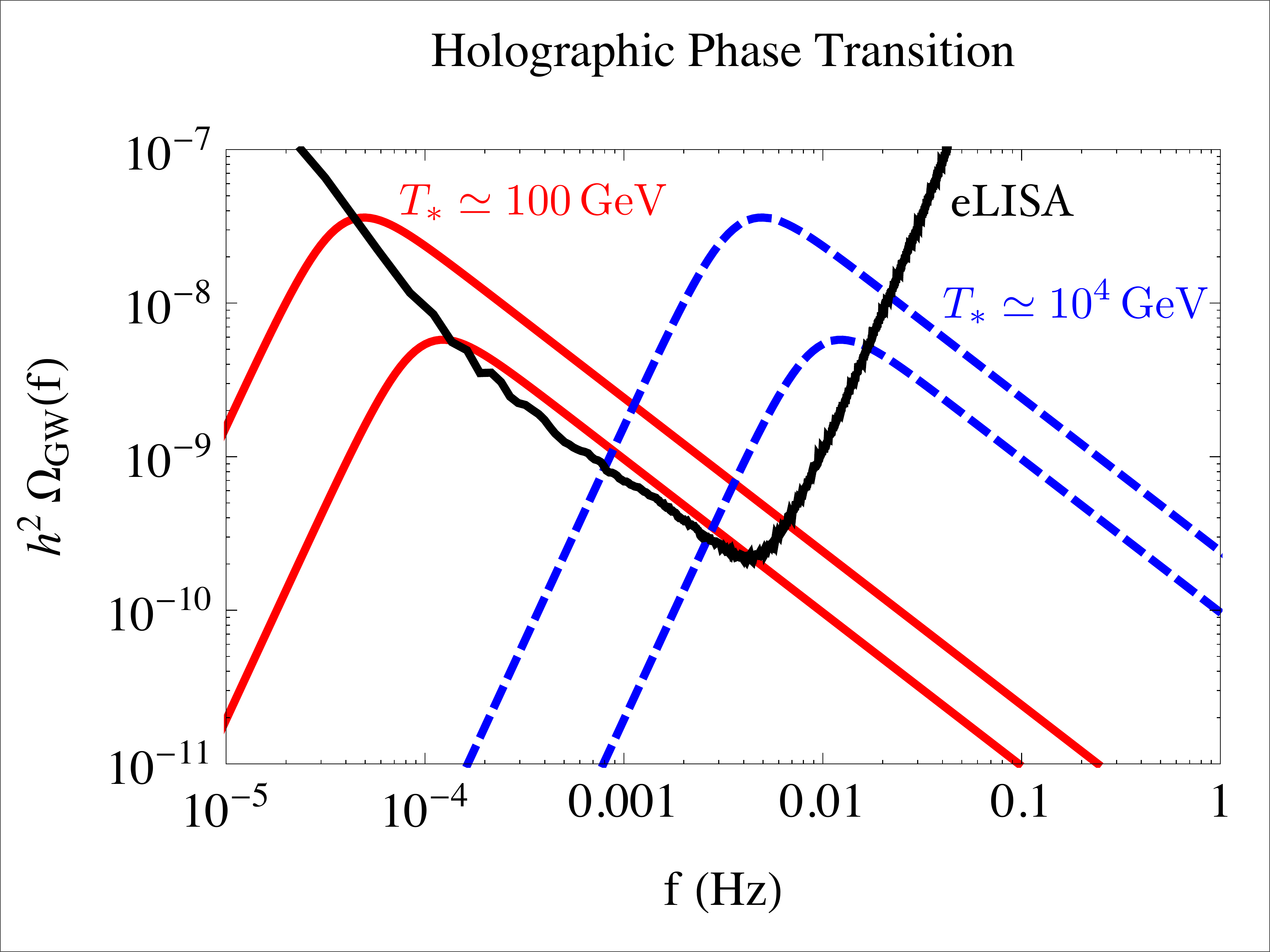}
\caption{\label{fig:PT} Taken from Ref.~\cite{elisapaper}. GW spectra in a model of first order PT connected to the radion stabilization in the Randal Sundrum model \cite{conformal}, for different values of the PT temperature and duration (for details, see \cite{elisapaper}). The signal amplitude is high since in this particular model the phase transition is strongly first order, and the bubble walls move at the speed of light.}
\end{minipage} 
\end{figure}

Two fundamental parameters enter in the GW spectrum, that depend on the underlying high-energy physics model: the string tension and the reconnection probability. The dimensionless parameter $G\mu$ representing the tension must be sufficiently small to satisfy observational constraints imposed by the CMB \cite{Ade:2013xla}. The reconnection probability is usually equal to one for cosmic strings in field theories, but it can be smaller for cosmic strings in string theory. The GW background from cosmic strings depends also on the typical size of the loops when they are produced: recently, different groups performing simulations reached an agreement on the loop size distribution, concluding that loops of all sizes are present, but the distribution is dominated by those produced within a few orders of magnitude of the horizon size \cite{christophe,Blanco-pillado}. Figure \ref{fig:strings}, taken from \cite{elisapaper}, shows the GW background for large loops for different values of $G\mu$ and of the reconnection probability. 

$\bullet$ {\bf First order phase transitions}: in the course of its adiabatic expansion, the universe might have undergone several PTs driven by the temperature decrease. Their nature depends on the particle theory model, but if they are first order they proceed through the nucleation and collision of broken phase bubbles, which lead to anisotropic stresses and therefore to the generation of GW (for a review, see \cite{elisapaper} and references therein). The EWPT in the standard model is a crossover, and it is not expected to lead to any appreciable cosmological signal. However, theories beyond the standard model can lead to a first order EWPT, giving a very rich phenomenology: besides a background of GW, a first order EWPT could provide Dark Matter candidates and baryogenesis. The recent discovery of the Higgs boson at the LHC confirms the paradigm of a scalar field-driven symmetry breaking in the early universe. There is yet no indication of new physics near the EW scale, but the order of the EWPT is not directly constrained by LHC data: several models leading to a first order EWPT remain viable, mainly based on expanding the Higgs sector with additional scalar states invisible at the energies probed by the LHC up to now (see e.g. \cite{beyondSM}). If the EWPT is sufficiently strongly first order, it could lead to a GW signal detectable by the space-based interferometer eLISA. Similarly, the QCDPT is also predicted to be a crossover by lattice simulations but it can become first order if the neutrino chemical potential is sufficiently large \cite{Schwarz:2009ii}. GW detection would help to probe the nature of these PTs, and provide interesting information on the underlying particle theory. Towards the end of a first order PT, the true vacuum bubbles collide and convert the entire universe to the broken phase. The collisions break the spherical symmetry of the bubble walls, generating a non-zero anisotropic stress \cite{astro-ph/9211004}. Moreover, they give rise to compression waves \cite{hind,Giblin:2014qia} and MHD turbulence (see e.g. \cite{turb}) in the surrounding fluid: their anisotropic stresses can act as a source of GW even after the merging of the bubbles is completed. The characteristic frequency of the GW spectrum is determined by the average size of the bubble towards the end of the PT, and its amplitude by the amount of tensor-type stress energy that is available: both these factors are strongly related to the strength of the first order PT. Figure \ref{fig:PT}, taken from \cite{elisapaper}, shows the GW signal for one example of first order PT occurring around  the EW scale \cite{conformal}.

\section{Conclusions}

We have little information about the physics and the processes operating in the very early universe, in particular for energies comprised between the scale of inflation (accessible through the observation of CMB anisotropies and polarization) and the EW scale (accessible at the LHC). Many interesting processes could take place in this energy range, such as reheating, baryogenesis, dark matter production, phase transitions and their remnants. Their occurrence and their characteristics ultimately depend on the high energy physics model that describes the very early universe. As discussed above, some of these processes can be powerful sources of GW, and due to their small interaction rate, GW would afterwards propagate freely until today. The detection of a GW signal from the early universe could therefore provide us with very significant information on the status of the universe at high energies. The characteristic frequency of the GW maps the temperature/energy scale of their generation process. GW by vacuum fluctuations during slow roll inflation are not visible by the next generation interferometers or by PTA. However, we have presented some examples of processes which generate a stochastic background of GW directly from anisotropic stresses, and which are more promising for detection. Therefore, GW are a powerful mean to learn about the early universe and high energy physics: the detection is extremely difficult, but it leads to a great payoff.

\end{document}